\documentclass[iop]{emulateapj}
\usepackage{color}
\usepackage{threeparttable}
\usepackage{ulem}
\usepackage{lscape}

\newcommand\ha{H$\alpha$~}

\newcommand\oiii{[O~{\sc iii}]~}

\shorttitle{Bulge-forming galaxies}
\shortauthors{Tadaki et al.}

\begin{document}


\title{Bulge-forming galaxies with an extended rotating disk at $z\sim2$}


\author{Ken-ichi Tadaki\altaffilmark{1}, 
Reinhard Genzel\altaffilmark{1,2,3}, 
Tadayuki Kodama\altaffilmark{4,5}, 
Stijn Wuyts\altaffilmark{6}, 
Emily Wisnioski\altaffilmark{1}, 
Natascha M. F{\"o}rster Schreiber\altaffilmark{1}, 
Andreas Burkert\altaffilmark{7,1}, 
Philipp Lang\altaffilmark{1}, 
Linda J. Tacconi\altaffilmark{1}, 
Dieter Lutz\altaffilmark{1}, 
Sirio Belli\altaffilmark{1}, 
Richard I. Davies\altaffilmark{1}, 
Bunyo Hatsukade\altaffilmark{4}, 
Masao Hayashi\altaffilmark{4}, 
Rodrigo Herrera-Camus\altaffilmark{1}, 
Soh Ikarashi\altaffilmark{8}, 
Shigeki Inoue\altaffilmark{9,10}, 
Kotaro Kohno\altaffilmark{11,12}, 
Yusei Koyama\altaffilmark{13}, 
J. Trevor Mendel\altaffilmark{1,7}, 
Kouichiro Nakanishi\altaffilmark{4,5}, 
Rhythm Shimakawa\altaffilmark{5}, 
Tomoko L. Suzuki\altaffilmark{5}, 
Yoichi Tamura\altaffilmark{11}, 
Ichi Tanaka\altaffilmark{13}, 
Hannah $\ddot{\mathrm{U}}$bler\altaffilmark{1}, and 
Dave J. Wilman\altaffilmark{7,1}}


\affil{\altaffilmark{1} Max-Planck-Institut f{\"u}r extraterrestrische Physik (MPE), Giessenbachstr., D-85748 Garching, Germany; tadaki@mpe.mpg.de}
\affil{\altaffilmark{2} Department of Physics, Le Conte Hall, University of California, Berkeley, CA 94720, USA}
\affil{\altaffilmark{3} Department of Astronomy, Hearst Field Annex, University of California, Berkeley, CA 94720, USA}
\affil{\altaffilmark{4} National Astronomical Observatory of Japan, 2-21-1 Osawa, Mitaka, Tokyo 181-8588, Japan}
\affil{\altaffilmark{5} Department of Astronomical Science, SOKENDAI (The Graduate University for Advanced Studies), Mitaka, Tokyo 181-8588, Japan}
\affil{\altaffilmark{6} Department of Physics, University of Bath, Claverton Down, Bath, BA2 7AY, UK}
\affil{\altaffilmark{7} Universit{\"a}ts-Sternwarte Ludwig-Maximilians-Universit{\"a}t (USM), Scheinerstr. 1, M{\"u}nchen, D-81679, Germany}
\affil{\altaffilmark{8} Kapteyn Astronomical Institute, University of Groningen, P.O. Box 800, 9700AV Groningen, The Netherlands}
\affil{\altaffilmark{9} Kavli Institute for the Physics and Mathematics of the Universe (WPI), UTIAS, The University of Tokyo, Chiba 277-8583, Japan}
\affil{\altaffilmark{10} Department of Physics, The University of Tokyo, 7-3-1 Hongo, Bunkyo, Tokyo 113-0033, Japan}
\affil{\altaffilmark{11} Institute of Astronomy, The University of Tokyo, 2-21-1 Osawa, Mitaka, Tokyo 181-0015, Japan}
\affil{\altaffilmark{12} Research Center for the Early Universe, The University of Tokyo, 7-3-1 Hongo, Bunkyo, Tokyo 113-0033, Japan}
\affil{\altaffilmark{13} Subaru Telescope, National Astronomical Observatory of Japan, 650 North A'ohoku Place, Hilo, HI 96720, USA}



\begin{abstract}

We present 0\arcsec.2-resolution Atacama Large Millimeter/submillimeter Array observations at 870 $\mu$m for 25 H$\alpha$-seleced star-forming galaxies around the main-sequence at $z=2.2-2.5$.
We detect significant 870 $\mu$m continuum emission in 16 (64\%) of these galaxies.
The high-resolution maps reveal that the dust emission is mostly radiated from a single region close to the galaxy center.
Exploiting the visibility data taken over a wide $uv$ distance range, we measure the half-light radii of the rest-frame far-infrared emission for the best sample of 12 massive galaxies with log$(M_*/M_\odot)>$11.
We find nine galaxies to be associated with extremely compact dust emission with $R_{1/2,870\mu\mathrm{m}}<1.5$ kpc, which is more than a factor of 2 smaller than their rest-optical sizes, $\langle R_{1/2,1.6\mu\mathrm{m}}\rangle$=3.2 kpc, and is comparable with optical sizes of massive quiescent galaxies at similar redshifts.
As they have an exponential disk with S$\acute{\mathrm{e}}$rsic index of $\langle n_{1.6\mu\mathrm{m}}\rangle$=1.2 in the rest-optical, they are likely to be in the transition phase from extended disks to compact spheroids.
Given their high star formation rate surface densities within the central 1 kpc of $\langle \Sigma \mathrm{SFR}_\mathrm{1kpc}\rangle=40~M_\odot$yr$^{-1}$kpc$^{-2}$, 
the intense circumnuclear starbursts can rapidly build up a central bulge with $\Sigma M_{*,\mathrm{1kpc}}>10^{10}~M_\odot$kpc$^{-2}$ in several hundred Myr, i.e. by $z\sim2$. 
Moreover, ionized gas kinematics reveal that they are rotation-supported with an angular momentum as large as that of typical star-forming galaxies at $z=1-3$. 
Our results suggest bulges are commonly formed in extended rotating disks by internal processes, not involving major mergers.
\end{abstract}


\keywords{galaxies: evolution --- galaxies: high-redshift --- galaxies: ISM}



\section{Introduction}


In the current paradigm of galaxy evolution, galaxies grow mainly by internal star formation along a fairly tight relationship between stellar mass and star formation (so-called main sequence), at a rate that is set by the balance between gas accretion from the cosmic web, internal star formation and outflows driven by active galactic nuclei (AGN), supernovae, and massive stars \citep{2010ApJ...718.1001B,2012MNRAS.421...98D,2013ApJ...772..119L}. 
Once galaxy masses reach the Schechter mass, $\log(M_*/M_\odot)\sim10.9$ \citep[e.g.,][]{2009ApJ...701.1765M,2013ApJ...777...18M,2013A&A...556A..55I}, star formation appears to drop within a short timescale of $\sim$1 Gyr \citep{2013ApJ...770L..39W,2015ApJ...804L...4M,2015ApJ...799..206B,2015ApJ...808..161O} and galaxies transition to the passive population below the main sequence. 

Star-forming galaxies on the main sequence have exponential optical light and mass distributions \citep[e.g.,][]{2011ApJ...742...96W,2015ApJ...811L..12W} with orbital motions dominated by rotation in $\sim$70\% of the massive star-forming galaxy population \citep[e.g.,][]{2009ApJ...706.1364F,2009ApJ...697.2057L,2012MNRAS.426..935S,2015ApJ...799..209W,2016MNRAS.457.1888S,2016ApJ...819...80P}. 
However, high-redshift star-forming galaxies exhibit significant random motions (turbulent) such that the disks are hot and geometrically thick \citep{2015ApJ...799..209W,2014ApJ...792L...6V}. 
In contrast, quiescent galaxies are more compact and cuspy than the star-forming galaxies at a given mass, at all redshifts \citep{2014ApJ...788...28V,2012ApJ...753..167B,2014ApJ...788...11L}.
Given these findings, quenching of star formation must be accompanied by significant structural change, from extended exponential distributions to more compact and more cuspy ones.

To explain the morphological transformation, two main evolutionary paths have been proposed in the literature. 
A slow cosmological path naturally follows from the strong redshift evolution of galaxy sizes, $R\propto(1+z)^{-1}$ \citep{2012ApJ...746..162N,2012ApJ...756L..12M,2014ApJ...788...28V,2015ApJS..219...15S}. 
Star-forming galaxies quench star formation and add to the passive population with approximately the same size in a later epoch \citep{2015ApJ...813...23V,2016arXiv160406459L}.
A second, fast path involves a downward transition in the mass--size plane, at approximately constant redshift \citep{2013ApJ...765..104B,2014ApJ...791...52B,2014MNRAS.438.1870D,2015MNRAS.450.2327Z}. 
This process requires a substantial ``compaction'' of the formally extended star-forming galaxies. 
One possible mechanism would be a major merger, which is known from observations and simulations to lead to substantial angular momentum redistribution, orbit reconfiguration and mixing \citep{1996ApJ...464..641M,2010ApJ...722.1666W}. 
Another possibility is an internal angular momentum redistribution within the star-forming disk. 
This process has been considered to be effective at high redshift \citep{1999ApJ...514...77N,2004A&A...413..547I,2004ApJ...611...20I,2008ApJ...688...67E,2008ApJ...687...59G,2011ApJ...741L..33B}, when galaxies are gas rich \citep{2013ApJ...768...74T} and effective viscous dissipation leads to radial inward transport of gas and stars with a time scale of a few 100 Myr \citep{2009ApJ...703..785D} and buildup of a central dense core (bulge component) through circumnuclear concentration of gas. 
\cite{2016ApJ...817L...9N} find in massive galaxies at $z\sim1.4$ that central 1 kpc regions are highly attenuated by dust and are responsible for half of the total star formation rate (SFR).
In conjunction with morphological quenching \citep{2009ApJ...707..250M,2014ApJ...785...75G}, and powerful AGN outflows \citep{2006MNRAS.365...11C,2006MNRAS.370..645B,2014ApJ...787...38F,2014ApJ...796....7G}, 
the compaction process may then lead to an inside-out quenching near the Schechter mass \citep{2015Sci...348..314T,2016MNRAS.458..242T}.

In the following paper, we report observations of submillimeter dust continuum emission with the Atacama Large Millimeter/submillimeter Array (ALMA) to search for compact concentrations of interstellar medium as a unique telltale sign of the fast evolutionary path.
An advantage of our study is there is no selection bias in galaxy morphologies.
Therefore, the key goal is to address the issue of morphological transformation from extended exponential disks to quiescent spheroids using the high-resolution ALMA/870 $\mu$m maps.
We show that bulges can be formed in massive extended, rotating disks at $z\sim2$, in a short timescale of several hundred Myr (Section \ref{sec;bulge}).

We assume a Chabrier initial mass function (IMF; \citealt{2003PASP..115..763C}) and adopt cosmological parameters of $H_0$ =70 km s$^{-1}$ Mpc$^{-1}$, $\Omega_{\rm M}$=0.3, and $\Omega_\Lambda$ =0.7.

\section{High-resolution 870 $\mu$m imaging}

\subsection{Sample selection}
\label{sec;sample}

Our sample is selected from a narrow-band imaging survey with the MOIRCS on the Subaru Telescope, tracing H$\alpha$ emission at $z=2.19\pm0.02$ or $2.53\pm0.02$ \citep{2013ApJ...778..114T, 2013IAUS..295...74K}, in the SXDF-UDS-CANDELS field, where 0\arcsec.18-resolution HST images at four passbands ($V_{606}$, $I_{814}$, $J_{125}$, and $H_{160}$) are publicly available \citep{2011ApJS..197...35G,2011ApJS..197...36K}.
The limiting H$\alpha$ line fluxes for the narrow-band survey correspond to dust-uncorrected SFRs of 4 $M_\odot$yr$^{-1}$ at $z=2.19$ and 10 $M_\odot$yr$^{-1}$ at $z=2.53$ \citep{1998ARA&A..36..189K}.
Interlopers with a different emission line such as \oiii at $z\sim3$ are excluded by utilizing colors to pick up the Balmer/4000\AA~break \citep{2015ApJ...806..208S}.
Follow-up spectroscopic observations demonstrate our method robustly picks up only galaxies at the redshift range of interest \citep{2011PASJ...63S.437T, 2013ApJ...778..114T}.
For ALMA observations of 25 galaxies, we prioritize bright objects in MIPS 24 $\mu$m maps, which are taken from the SpUDS Spitzer Legacy program (PI: James Dunlop), to increase the feasibility of detection in the ALMA Early Science phase. Four out of 25 galaxies are not detected at 24 $\mu$m.

\subsection{Galaxy properties}
\label{sec;properties}

To derive galaxy properties, we use the 3D-HST catalog, including photometric data at 18 bands from $U-$band to 8.0 $\mu$m \citep{2014ApJS..214...24S,2016ApJS..225...27M}.
Using the {\tt FAST} code \citep{2009ApJ...700..221K}, we perform spectral energy distribution (SED) fitting with stellar population synthesis models of \citet{2003MNRAS.344.1000B} under a solar metallicity, exponentially declining star formation histories (SFHs), and dust attenuation law of \citet{2000ApJ...533..682C} to estimate stellar masses.
We also create a deep PACS 160 $\mu$m map from archival data with {\tt UNIMAP} \citep{2015MNRAS.447.1471P} and extract sources on the basis of 24 $\mu$m priors (see also \citealt{2011A&A...532A..90L} for the methodology).
Following the recipes of \citet{2011ApJ...738..106W}, we compute total SFRs from a combination of the rest-frame 2800 \AA~and infrared luminosities with PACS 160 $\mu$m or MIPS 24 $\mu$m fluxes ($L_\mathrm{IR}$).
For four galaxies without detection at mid-infrared, we use H$\alpha$-based SFRs with dust correction from SED modeling \citep{2015ApJ...811L...3T}. 
Table \ref{tab;1} summarizes the galaxy properties for our ALMA sample of 25 galaxies.
We adopt uncertainties of $\pm$0.15 dex for the stellar mass and $\pm$0.20 dex for the SFR taking into account systematic errors although uncertainties associated with photometry measurements are somewhat smaller \citep{2011ApJ...738..106W}.
For dusty star-forming galaxies such as submillimeter sources, the random uncertainties in the stellar mass estimates could be larger because the stellar components hide behind dust.

\begin{figure*}[t]
\begin{center}
\includegraphics[scale=1.1]{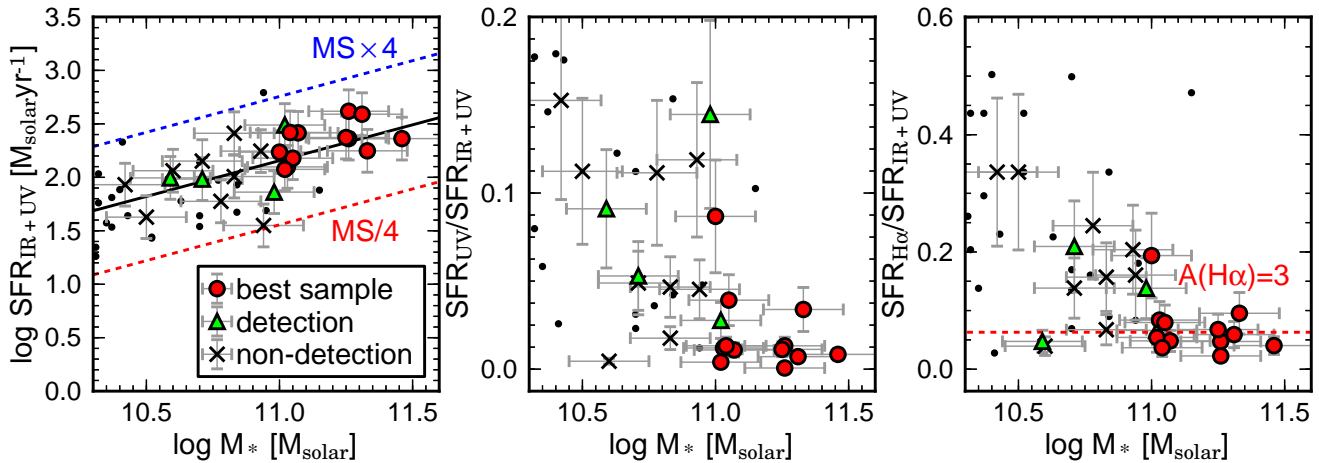}
\caption{
(left) Stellar mass versus star formation rate for our ALMA sample of 25 star-forming galaxies at $z=2.2$ or $z=2.5$. 
Red circles indicate the best sample that is detected both in the low-resolution and high-resolution 870 $\mu$m maps and green triangles show all objects detected in either maps. 
Small dots show our parent sample of galaxies identified by the narrow-band H$\alpha$ imaging. They lie on/around the main-sequence of star formation at $z=2.0-2.5$ (solid line; \citealt{2014ApJ...795..104W}). 
(middle) Ratio of UV-based SFR over total one, derived from UV and infrared luminosities, as a function of stellar mass. (right) Ratio of H$\alpha$-based SFR over total one. H$\alpha$ fluxes are measured in the narrow-band maps \citep{2013ApJ...778..114T}. A dashed red line corresponds to a dust extinction of $A_{\mathrm{H}\alpha}=3$ mag.
}
\label{fig;MS}
\end{center}
\end{figure*}

SFRs of galaxies are well correlated with their stellar masses, with a scatter of $\pm$0.3 dex \citep[e.g.,][]{2007ApJ...660L..43N,2007ApJ...670..156D,2007A&A...468...33E,2009ApJ...698L.116P,2011ApJ...739L..40R,2013ApJ...777L...8K,2012ApJ...754L..29W,2014ApJ...795..104W,2015ApJ...815...98S,2015A&A...581A..54T}.
Our ALMA sample of 25 galaxies is on/around the star-formation main sequence (Figure \ref{fig;MS}), indicating that they probe the normal star-forming population at $z\sim2$.


At $z=2.2-2.5$, HST/WFC3 $H_{160}$-band traces the rest-optical light ($\lambda_{rest}=0.46-0.50~\mu$m) of galaxies. 
The structural parameters such as circularized half-light radius and S$\acute{\mathrm{e}}$rsic index are derived with {\tt GALFIT} \citep{2010AJ....139.2097P} in the $H_{160}$-band maps \citep{2012ApJS..203...24V, 2014ApJ...788...28V}. 
We do not use U4-27289 and U4-16795 for optical size arguments because the best-fit S$\acute{\mathrm{e}}$rsic index reached the constrained limit ($n=8.0$ or $n=0.2$).


\begin{table*}[t]
\begin{threeparttable}
\caption{Galaxy properties for our ALMA sample of 25 star-forming galaxies. \label{tab;1}}
\begin{tabular}{llccccccccc}
\hline
3D-HST ID & $z_\mathrm{NB}$\tablenotemark{a} & $\log~M_*$\tablenotemark{b}& $\log~$SFR\tablenotemark{b} & SNR$_\mathrm{0.5}$\tablenotemark{c} & SNR$_\mathrm{0.2}$\tablenotemark{c} & $S_\mathrm{aper}$\tablenotemark{c} & $S_\mathrm{model}$\tablenotemark{d} & $R_\mathrm{1/2}$\tablenotemark{d} & $R_\mathrm{1/2,cor}$\tablenotemark{e} & $v_\mathrm{rot}/\sigma_0$\tablenotemark{f}\\
(Skelton+14) & & ($M_\odot$) & ($M_\odot$yr$^{-1}$) & & & (mJy) & (mJy) & (arcsec) & (arcsec) & \\
\hline
U4-13952 & 2.19 & 11.33 & 2.25 &13.4      & 7.9       & 2.51$\pm$0.31 & 2.94$\pm$0.55 & 0.24$\pm$0.04 & 0.28$\pm$0.06 & 3.8$\pm$1.3\\  
U4-34817 & 2.19 & 11.26 & 2.36 &  7.8      & 5.4       & 1.73$\pm$0.28 & 2.13$\pm$0.78 & 0.31$\pm$0.10 & 0.38$\pm$0.12 & \ha detection\\  
U4-20704 & 2.19 & 11.46 & 2.36 &  8.1      & 6.3       & 3.00$\pm$0.40 & 4.28$\pm$1.11 & 0.44$\pm$0.10 & 0.48$\pm$0.11 & 4.2$\pm$1.4\\
U4-28702 & 2.19   & 11.03 & 2.10 &10.1      & 9.7       & 1.73$\pm$0.36 & 1.64$\pm$0.31 & 0.10$\pm$0.02 & 0.13$\pm$0.03 & \\
U4-36568 & 2.19 & 11.02 & 2.49 &  4.0      & $<$5.0 & 0.71$\pm$0.24 & & & & 5.3$\pm$1.8\\   
U4-24247 & 2.19 & 10.71 & 1.98 &  4.4      & $<$5.0 & 1.09$\pm$0.36 & & & & \ha detection\\
U4-32171 & 2.19   & 10.71 & 2.15 & $<$4.0 & $<$5.0 &  & & & &\\  
U4-11582 & 2.19 & 10.83 & 2.01 & $<$4.0 & $<$5.0 &  & & & & 6.9$\pm$2.4\\  
U4-27289 & 2.19   & 10.78 & 1.78 & $<$4.0 & $<$5.0 &  & & & &\\
U4-36247 & 2.19 & 11.07 & 2.42 & 13.5     & 16.0     & 1.80$\pm$0.24 & 1.41$\pm$0.18 & 0.05$\pm$0.01& 0.07$\pm$0.02 & 3.5$\pm$2.3\\  
U4-32351 & 2.19 & 11.05 & 2.18 & 6.5       & 6.8       & 0.95$\pm$0.26 & 0.74$\pm$0.24 & 0.10$\pm$0.04& 0.17$\pm$0.08 & 5.2$\pm$0.9\\  
U4-18807 & 2.19 & 10.98 & 1.86 & $<$4.0 & 5.5       & 0.58$\pm$0.26 & & &  & 7.1$\pm$4.9\\
U4-27939 & 2.19   & 10.60 & 2.06 & $<$4.0 & $<$5.0 & & & & &\\
U4-14574 & 2.19   & 10.59 & 1.99 & 4.0       & $<$5.0 & 1.20$\pm$0.46 &  & & & \\
U4-15198 & 2.53   & 10.93 & 2.24 & $<$4.0 & $<$5.0 & &  & & & \\
U4-16795 & 2.53   & 11.26 & 2.62 & 31.0     & 29.2     & 4.59$\pm$0.31 & 4.46$\pm$0.27 & 0.12$\pm$0.01 & 0.13$\pm$0.01\\  
U4-34138 & 2.53 & 11.00 & 2.24 & 9.7       & 11.4     & 1.60$\pm$0.29 & 1.10$\pm$0.19 & 0.06$\pm$0.02 & 0.08$\pm$0.03 & 3.8$\pm$2.0\\  
U4-28473 & 2.53 & 11.31 & 2.59 & 26.0     & 22.5     & 4.87$\pm$0.45 & 5.12$\pm$0.39 & 0.13$\pm$0.01 & 0.14$\pm$0.02 & 6.1$\pm$4.0\\
U4-33135 & 2.53   & 11.02 & 2.07 & 8.6       & 9.8       & 1.47$\pm$0.34 & 1.27$\pm$0.25 & 0.07$\pm$0.02 & 0.09$\pm$0.03 & \\
U4-27046 & 2.53 & 10.83 & 2.41 & $<$4.0 & $<$5.0 & & & & & \ha detection\\
U4-16504 & 2.53   & 11.25 & 2.37 & 20.4     & 15.7     & 2.82$\pm$0.23 & 3.16$\pm$0.34 & 0.15$\pm$0.02 & 0.17$\pm$0.03 &\\  
U4-11780 & 2.53   & 10.42 & 1.93 & $<$4.0 & $<$5.0 &  & & \\
U4-13197 & 2.53   & 10.94 & 1.55 & $<$4.0 & $<$5.0 &  &  & & &\\
U4-34617 & 2.53   & 11.04 & 2.42 & 10.6     & 13.0     & 1.67$\pm$0.28 & 0.93$\pm$0.13 &  0.02$\pm$0.01 & 0.04$\pm$0.02 &\\  
U4-14870 & 2.53   & 10.50 & 1.63 & $<$4.0 & $<$5.0 &  & & & & \\ 
\hline
\end{tabular}
\begin{tablenotes}
\item[a] Redshifts derived from the narrow-band imaging survey with Subaru \citep{2013ApJ...778..114T}.
\item[b] Stellar masses estimated with SED modeling and total star formation rates computed from rest-frame 2800 \AA and infrared luminosities \citep{2011ApJ...738..106W}. We adopt uncertainties of $\pm$0.15 dex for the stellar mass and $\pm$0.20 dex for the SFR.
\item[c] Signal-to-noise ratios of the peaks in 0\arcsec.5- and 0\arcsec.2-resolution ALMA/870 $\mu$m maps. We measure total fluxes, $S_\mathrm{aper}$, with 1\arcsec.5 aperture in the 0\arcsec.5-resolution maps or with 1\arcsec.0 aperture in the 0\arcsec.2-resolution maps.
\item[d] 870 $\mu$m fluxes and half-light radii for the best-fit exponential model. 
\item[e] Half-light radii corrected for residual emission with $S_\mathrm{extra}=$0.4 mJy (section \ref{sec;size_measurements}).
\item[f] Ratios of rotation velocity to local velocity dispersion measured with KMOS.
\end{tablenotes}
\end{threeparttable}
\end{table*}

\begin{figure}[h]
\begin{center}
\includegraphics[scale=1.1]{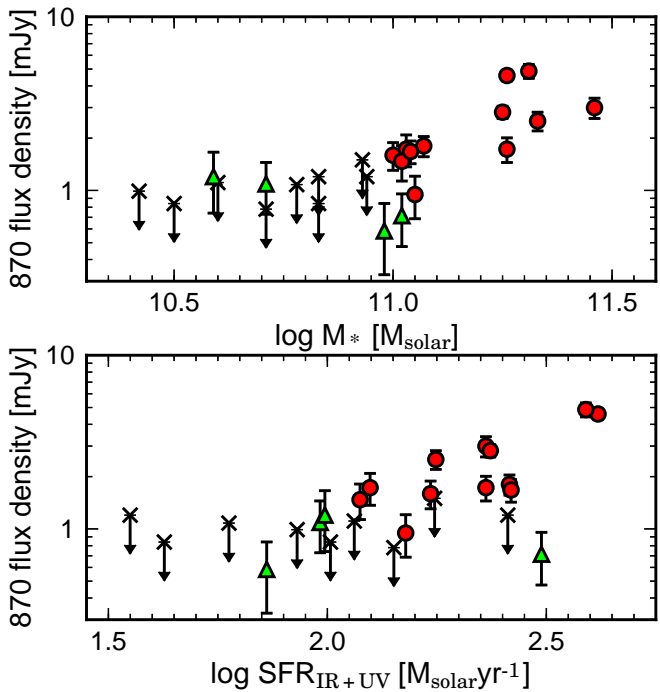}
\caption{
Comparisons between 870 $\mu$m fluxes and galaxy properties. Symbols are the same as in Figure \ref{fig;MS}. For non-detected objects, the 3$\sigma$ upper limits are plotted.
}
\label{fig;870flux}
\end{center}
\end{figure}

\subsection{ALMA observations}
\label{sec;observation}

We have carried out ALMA observations for 25 galaxies on the main-sequence at $z=2$ with 32--49 antennas and baseline lengths of 20--1600 m.
On-source time is 6--8 minutes per object.
We use the band 7 receivers with the 64-input correlator in Time Division Mode in a central frequency of 345 or 350 GHz ($\sim$870 $\mu$m).
We utilize the Common Astronomy Software Application package ({\tt CASA}; \citealt{2007ASPC..376..127M}) for the data calibration.
We reconstruct two kinds of clean maps: low-resolution maps with $uv$-taper of the on-sky FWHM=0\arcsec.5 and high-resolution ones with natural weighting.
The synthesized beamsizes are 0\arcsec.47-0\arcsec.54 and 0\arcsec.15-0\arcsec.21, respectively.
We measure total fluxes, $S_\mathrm{aper}$, with 1\arcsec.5 aperture photometry in the low-resolution maps or with 1\arcsec.0 aperture in the high-resolution maps.
Uncertainties of total fluxes are derived by computing standard deviations of 50 random apertures in each of the maps.
The rms levels are 98-142 $\mu$Jy beam$^{-1}$ for the low-resolution maps and 56-74 $\mu$Jy beam$^{-1}$ for the high-resolution maps

For detections, we adopt a 4$\sigma$ threshold in a peak flux density on the low-resolution maps or 5$\sigma$ on the high-resolution maps, where sources with negative signal become zero.
We have detected 16 out of the 25 galaxies either in the low-resolution or the high-resolution maps.
Massive and active star-forming galaxies tend to be bright at 870 $\mu$m (Figure \ref{fig;MS}). 
For galaxies at at similar redshifts ($z$=2.19 or 2.53), we find the measured 870 $\mu$m fluxes to be correlated both with stellar masses and SFRs (Figure \ref{fig;870flux}).
The Pearson product-moment correlation coefficients are 0.66 for stellar masses and 0.69 for SFRs.
The detection rate is 100\% (13/13) in the stellar mass range of log($M_*$/$M_\odot$)$>11$ while some galaxies with high SFRs are not detected.
Given the correlation and the mass dependence of the detection rate, stellar masses are likely to be a good predictor of 870 $\mu$m fluxes \citep{2016arXiv160600227D}.
The total average flux is $\langle S_\mathrm{aper}\rangle$=2.0 mJy (0.6-4.9) in all detected objects, fainter than those of classical submillimetre galaxies identified by single dish telescopes \citep[e.g.,][]{2015ApJ...799...81S}.

\subsection{KMOS observations}
\label{sec;kmos}

We have observed 12 of 25 galaxies with the near-infrared integral-field spectrometer KMOS on the Very Large Telescope (VLT) as part of the KMOS$^\mathrm{3D}$ survey \citep{2015ApJ...799..209W} to study the spatially resolved ionized gas kinematics of these sources.
For our ALMA sample, a typical integration time is 11 hours.
We reduced the data with the Software Package for Astronomical Reduction ({\tt SPARK}; \citealt{2013A&A...558A..56D}).
All of our targets show \ha emission and are spectroscopically confirmed to be at $z=2.19$ or $z=2.53$ within the expected uncertainty from the width of the narrow-band filters ($\Delta z=\pm0.02$).

Our method to derive kinematic parameters is described in detail by \cite{2016ApJ...826..214B} (see also \citealt{2015ApJ...799..209W,2016arXiv160303432W}). 
Here, we summarize procedures to obtain the rotation velocity ($v_\mathrm{rot}$), the local velocity dispersion ($\sigma_0$), and the disk angular momentum ($j_\mathrm{disk}$) from the reduced 3D cubes. 
A fundamental assumption is that high-redshift star-forming galaxies are symmetric oblate, thick disks with an exponential profile, which is supported by observations \citep{2009ApJ...697.2057L, 2009ApJ...706.1364F,2011ApJ...742...96W,2014ApJ...785...75G,2014ApJ...792L...6V}. 
First, we create velocity field and velocity dispersion maps by fitting Gaussian profiles to the data in each spatial pixel.
After determining the largest total velocity gradient and the radius at which this velocity gradient reaches a maximum value ($R_\mathrm{max}$), we measure rotation velocities at $R_\mathrm{max}$ and local velocity dispersions in outer disks. 
Here we correct for observational effects (inclination and beam smearing) on the basis of structural parameters for the rest-optical light in the H$_{160}-$band maps. 
For symmetric oblate disks, the inclination, $i$, is estimated from the projected minor-to-major axis ratio, $q_\mathrm{obs}=b/a$, as $\sin^2(i)=(1-q_\mathrm{obs}^2)/(1-q_\mathrm{int}^2)$,
with an intrinsic finite thickness of $q_\mathrm{int}=0.15-0.25$ \citep{2009ApJ...697.2057L, 2009ApJ...706.1364F, 2015ApJ...799..209W, 2016arXiv160303432W}. 
The impact of the beam smearing depends on the ratio of half-light radius to HWHM of the PSF, $R_{1/2}/R_\mathrm{PSF}$, and $R_\mathrm{max}/R_{1/2}$. 
We also correct for turbulent pressure to derive a circular velocity, $v_{circ}$, and the correction factor is 1.03--1.32 in our sample. 
The specific angular momentum of ionized gas is computed as 

\begin{equation}
j_\mathrm{disk}=k_\mathrm{disk}\times v_\mathrm{circ}\times R_{1/2}.
\end{equation}

\noindent
Here, we take into account deviations from exponential profiles.
The correction factors, $k_\mathrm{disk}$, are $k_\mathrm{disk}=1.19$ in $n=1$, $k_\mathrm{disk}=2.29$ in $n=4$ and $k_\mathrm{disk}=0.89-1.36$ in our sample \citep{2012ApJS..203...17R}. 

We eventually obtain the kinematic parameters for nine galaxies (Table \ref{tab;1}).
They are all rotation-supported with $\langle v_\mathrm{rot}/\sigma_0\rangle=5.1~(3.5-7.1)$ as is the case for most of galaxies on/around the main-sequence \citep{2015ApJ...799..209W}. 
Therefore, our ALMA sample is a typical star-forming population at $z\sim2$ in star-forming activity, morphology and kinematics.

\begin{figure*}
\begin{center}
\includegraphics[scale=1.1]{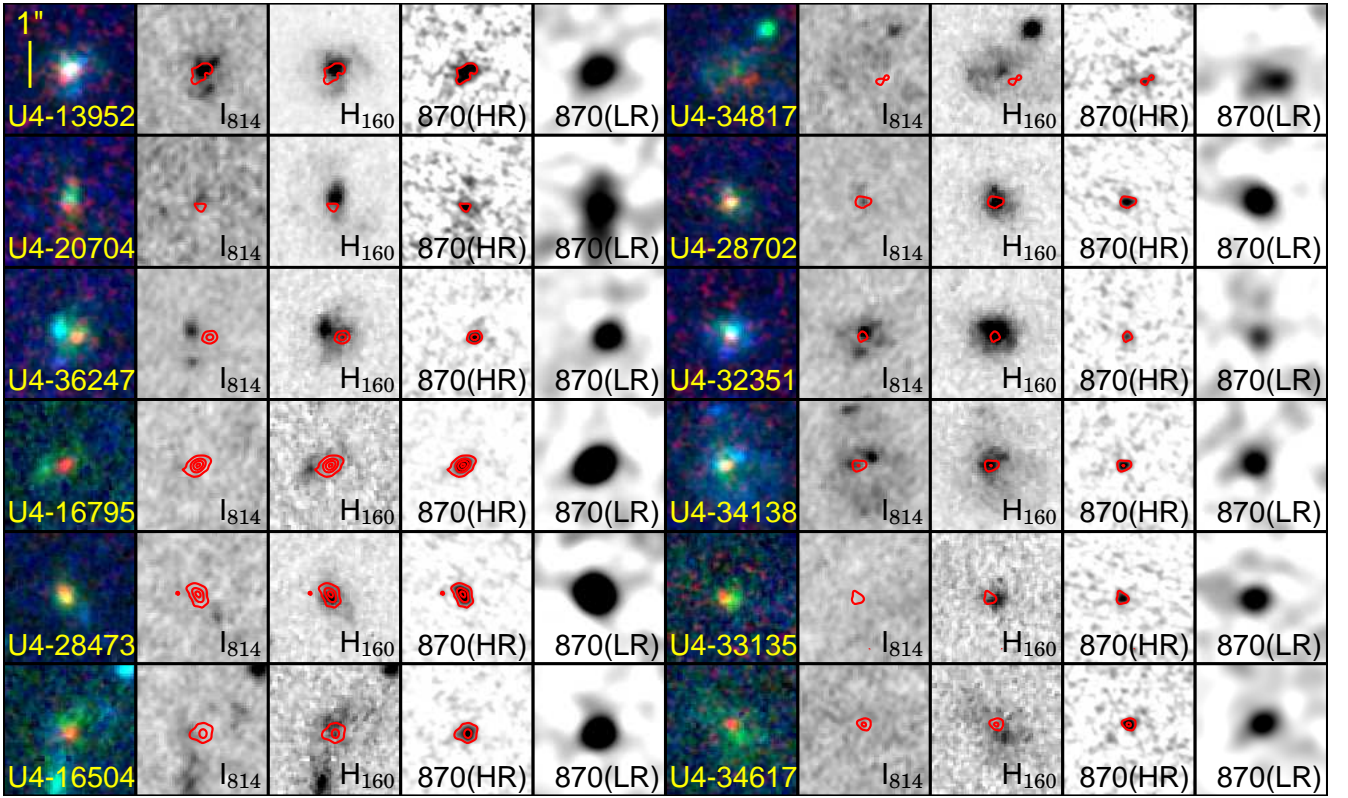}
\caption{Three-color images with HST/$I_{814}$, $H_{160}$ and ALMA/870 $\mu$m-band (3\arcsec $\times$3\arcsec) for our sample of 12 galaxies with 870 $\mu$m size measurements. 
Red contours display the 870 $\mu$m flux densities in the high-resolution maps and are plotted every 8$\sigma$, starting at 4$\sigma$. \label{fig;HST}}
\end{center}
\end{figure*}

\section{Spatial extent of star formation within galaxies}
\label{sec;size}

The most straightforward way to know the subsequent evolution of galaxy morphologies is to reveal where and how much stars are formed within galaxies at the observed epoch. 
Many previous studies use the rest-frame UV or H$\alpha$ maps to investigate the spatial distribution of star formation \citep[e.g.,][]{2011ApJ...733..101G,2012ApJ...747L..28N,2016ApJ...828...27N,2013ApJ...779..135W}. 
However, for our ALMA sample of massive galaxies, the measured ratios of SFR$_\mathrm{UV}$/SFR$_\mathrm{IR+UV}$ and SFR$_\mathrm{H\alpha}$/SFR$_\mathrm{IR+UV}$ indicate that $\sim$99\% of the total SFR is obscured by dust and even \ha emission misses 90-95\% of star formation, corresponding to a dust extinction of $A_\mathrm{H\alpha}\sim$3 mag (Figure \ref{fig;MS}). 
Therefore, the 870 $\mu$m maps tracing dust emission itself have a great advantage over H$\alpha$ to approximately provide the spatial distribution of star formation within galaxies if the dust temperature is constant across galaxies.
In this section, exploiting the ALMA data taken in the extended configuration, we study the spatial distributions of star formation within galaxies. 
We use the best sample of 12 galaxies which are detected both in low-resolution and high-resolution maps because the detections in a wide range of $uv$ distance allow us to constrain the spatial extent of dust continuum emission. 
Using the similar spatial resolution maps with HST/WFC3, we directly compare dusty star-forming regions with the rest-optical light mainly from stars.

\begin{figure*}
\begin{center}
\includegraphics[scale=0.9]{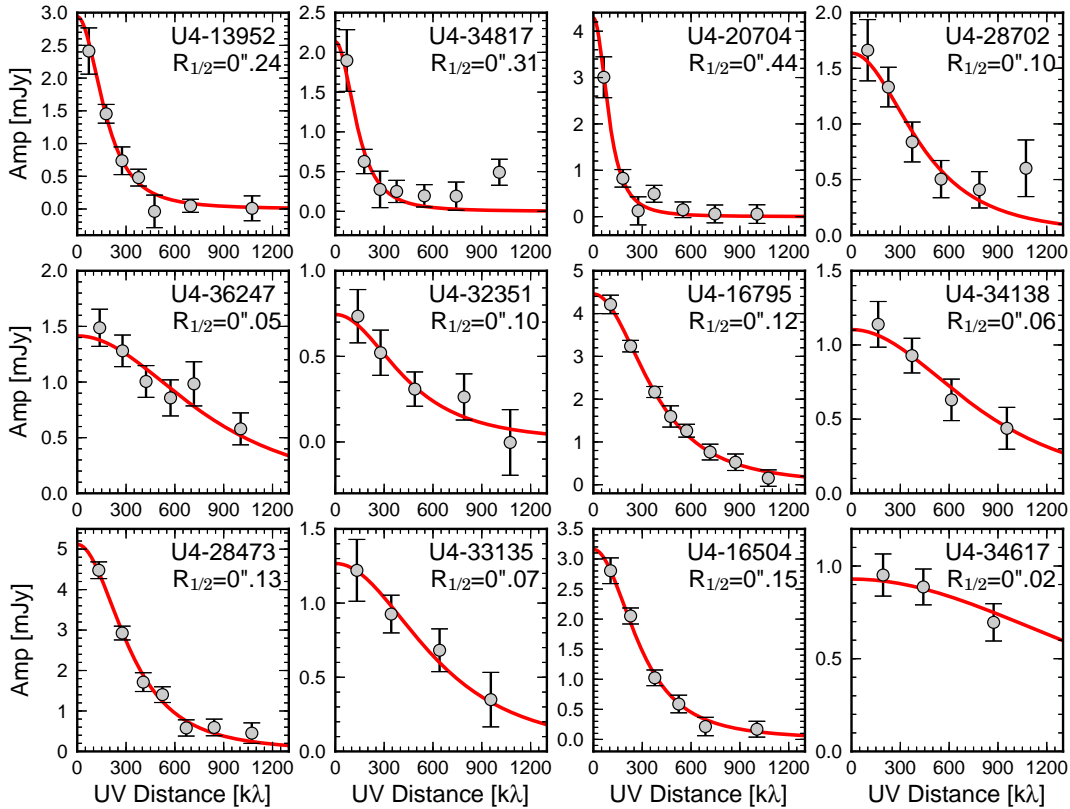}
\caption{Visibility amplitudes versus uv distances for our sample of 12 galaxies with size measurements. 
Red lines indicate the best-fitting model, $S_\mathrm{model}\times k_0^3/(u^2+k_0^2)^{3/2}$. 
The fitting was done with individual visibilities, not plotted in this figure. 
For reference, the amplitudes averaged over $uv$ distance are shown by gray circles.}
\label{fig;visibility}
\end{center}
\end{figure*}

\subsection{High-resolution 870 $\mu$m maps}

First, we visually inspect the high-resolution 870 $\mu$m maps before quantitatively measure sizes of the dust continuum emission. 
Figure \ref{fig;HST} shows the ALMA maps along with the similar resolution ACS/I$_{814}$ (rest-UV) and WFC3 $H_{160}$ (rest-optical) light distributions for the 12 galaxies. 
For about half of our sample, there is very little UV emission probably due to strong dust extinction. 
A common remarkable feature is that 870 $\mu$m emission is radiated from a single region close to the rest-optical center rather than multiple components like star-forming clumps in disks, seen in the rest-UV or H$\alpha$ maps.
Given that they are highly obscured, the concentrated component at 870 $\mu$m is primarily responsible for star formation in the galaxies.
An absence of dust emission in UV clumps means their 870 $\mu$m flux density could be below the lower limit of our ALMA observations. 
Also note that these high-resolution maps are sensitive to compact components with a spatial scale of $\sim$0\arcsec.2 and we might miss extended, diffuse components.
We assess impacts by faint components and/or extended ones in the next section.
For U4-34817 and U4-20704, the 870 $\mu$m emission appear to be faint in the high-resolution maps in spite of a relatively large flux ($S_{aper}=1.7, 3.0$ mJy).
They are likely to be associated with extended emission as they are more robustly detected in the low-resolution maps.

\subsection{Size measurements for 870 $\mu$m continuum emission}
\label{sec;size_measurements}

We measure half-light radii ($R_{1/2}$) of the primary component for dust emission, identified in the high-resolution maps. 
As interferometric telescopes do not directly provide images, the Fourier transform must be performed to reconstruct maps (clean algorithm). 
Then, image properties such as rms level, spatial resolution and source structures depend on clean parameters. 
To avoid these uncertainties, we perform visibility fitting with a circular exponential profile as seen in the rest-optical light. 
In previous studies, a Gaussian model is commonly used for size measurements in $u-v$ plane \citep{2015ApJ...810..133I,2015ApJ...799...81S,2015ApJ...811L...3T}. 
However, a radial profile of galaxy disks is approximately described by an exponential function, $n=1$ \citep[e.g.,][]{2011ApJ...742...96W}.
As our concern in this paper is primarily size differences between the rest-optical and 870 $\mu$m emission, an exponential model is preferred for a consistent comparison.

For an exponential function in the image plane, $f(R)=\exp(-1.678 R/R_{1/2})$, the Hankel transform (equivalent to a two-dimensional Fourier transform) is given by 

\begin{equation}
g(u)=S_\mathrm{model}\times\frac{k_0^3}{(u^2+k_0^2)^{3/2}},
\end{equation}

\begin{table*}
\begin{threeparttable}
\caption{Galaxy properties for 12 galaxies with 870 $\mu$m size measurements. \label{tab;2}}
\begin{tabular}{lccccccc}
\hline
3D-HST ID\ \ \ \  & \ \ \ $n_\mathrm{1.6\mu m}$\tablenotemark{a}\ \ \  & \ \ $R_\mathrm{1/2,1.6\mu m}$\tablenotemark{a}\ \  & \ \ $R_\mathrm{1/2,870\mu m}$\tablenotemark{b}\ \  & \ \ log$\Sigma M_{*\mathrm{1kpc}}$\tablenotemark{c}\ \ & \ \ log$\Sigma$SFR$_{\mathrm{1kpc}}$\tablenotemark{d}\ \  & \ \ \ \ \ log$\tau_\mathrm{bulge}$\tablenotemark{e}\ \ \ \ \  & \ \ log$\tau_\mathrm{depl}$\tablenotemark{f}\ \ \\
&  & (kpc) & (kpc) & $M_\odot$kpc$^{-2}$ & $M_\odot$yr$^{-1}$kpc$^{-2}$ & (yr) & (yr)\\
\hline
U4-13952 & 2.2$\pm$0.2 & 3.6$\pm$0.2 & 2.3$\pm$0.5 & 9.63$\pm$0.15  & 1.00$\pm$0.23  & 8.96$\pm$0.26  & 8.56$\pm$0.31 \\
U4-34817 & 0.6$\pm$0.6 & 5.0$\pm$0.5 & 3.1$\pm$1.0 & 9.17$\pm$0.15  & 0.93$\pm$0.30  & 9.14$\pm$0.30  & 8.48$\pm$0.31 \\
U4-20704 & 3.4$\pm$0.2 & 5.8$\pm$0.8 & 4.0$\pm$0.9 & 9.83$\pm$0.15  & 0.72$\pm$0.26  & 8.96$\pm$0.41  & 8.55$\pm$0.31 \\
U4-28702 & 1.2$\pm$0.5 & 2.5$\pm$0.3 & 1.0$\pm$0.3 & 9.45$\pm$0.15  & 1.28$\pm$0.22  & 8.79$\pm$0.23  & 8.52$\pm$0.31 \\
U4-36247 & 0.5$\pm$0.4 & 2.9$\pm$0.3 & 0.6$\pm$0.2 & 9.68$\pm$0.15  & 1.76$\pm$0.20  & 8.19$\pm$0.25  & 8.39$\pm$0.31 \\
U4-32351 & 1.9$\pm$0.8 & 2.6$\pm$0.2 & 1.4$\pm$0.6 & 9.56$\pm$0.15  & 1.28$\pm$0.24  & 8.74$\pm$0.26  & 8.49$\pm$0.31 \\
U4-16795 &                      &                      & 1.0$\pm$0.1 & 9.38$\pm$0.15  & 1.81$\pm$0.20  & 8.29$\pm$0.21  & 8.34$\pm$0.31 \\
U4-34138 & 1.2$\pm$0.2 & 5.8$\pm$0.4 & 0.6$\pm$0.2 & 9.41$\pm$0.15  & 1.55$\pm$0.21  & 8.55$\pm$0.21  & 8.41$\pm$0.31 \\
U4-28473 & 1.5$\pm$1.2 & 2.4$\pm$0.5 & 1.2$\pm$0.1 & 9.73$\pm$0.15  & 1.73$\pm$0.20  & 8.16$\pm$0.27  & 8.37$\pm$0.31 \\
U4-33135 & 1.0$\pm$2.1 & 1.5$\pm$0.8 & 0.8$\pm$0.2 & 9.76$\pm$0.15  & 1.36$\pm$0.21  & 8.50$\pm$0.29  & 8.49$\pm$0.31 \\
U4-16504 & 1.0$\pm$0.8 & 3.1$\pm$0.8 & 1.4$\pm$0.2 & 9.46$\pm$0.15  & 1.43$\pm$0.21  & 8.64$\pm$0.22  & 8.44$\pm$0.31 \\
U4-34617 & 0.9$\pm$0.3 & 5.0$\pm$0.7 & 0.3$\pm$0.2 & 9.17$\pm$0.15  & 1.76$\pm$0.20  & 8.40$\pm$0.20  & 8.35$\pm$0.31 \\
\hline
\end{tabular}
\begin{tablenotes}
\item[a] S$\acute{\mathrm{e}}$rsic indices and half-light radii at 1.6 $\mu$m. We do not use U4-16795 because the best-fit S$\acute{\mathrm{e}}$rsic index reaches the constrained limit of $n=8$.
\item[b] Half-light radii at 870 $\mu$m.
\item[c] Stellar mass surface density within a central 1 kpc calculated in stellar mass maps.
\item[d] SFR surface density within a central 1 kpc calculated from the best-fit exponential models at 870 $\mu$m and total SFRs.
\item[e] Bulge formation timescales to reach the stellar mass surface density of log($\Sigma M_{\mathrm{bulge}}$/$M_\odot$kpc$^{-2}$)=10 (Equation (\ref{eq;1})).
\item[f] Gas depletion timescales by star formation and outflows (Equation (\ref{eq;2})).
\end{tablenotes}
\end{threeparttable}
\end{table*}

\noindent
where $S_\mathrm{model}$ is the total flux of the model and $k_0$ is the spatial frequency to characterize a spatial extent. 
For the visibility fitting, we use the {\tt UVMULTIFIT} tool \citep{2014A&A...563A.136M}, which outputs full width at half maximum (FWHM) of a two-dimensional flux distribution (FWHM=0.826 $R_{1/2}$). 
In some cases, unexpected 870 $\mu$m sources are serendipitously detected within the primary beam. As they affect the visibility amplitudes of our main targets, we create a model of the interlopers and subtract it from the observed visibilities in advance. 
Figure \ref{fig;visibility} shows the observed visibility amplitudes after binning and the best-fit models, whose size and flux density are summarized in Table \ref{tab;1}. 
We obtain uncertainties in the sizes from fitting errors.
If adopting a circular Gaussian model, the estimated 870 $\mu$m sizes would become smaller by 7$\pm$6\%.

We also search for systematic positional offsets between ALMA/870 $\mu$m and HST/1.6 $\mu$m centers.
There is a small systematic offset of 19 mas in R.A. and 70 mas in declination.
U4-34817 has a significant offset of 405 mas between 870 $\mu$m and 1.6 $\mu$m peak.
Except for this galaxy, a mean separation is 130$\pm$68 mas, supporting the dust continuum emission arises from a central region of the galaxies.

For the size measurements, we investigate the impact of residual emission, which could be due to an additional extended component over entire disks, sub-structures like clumps, or deviations from an exponential model. 
In clean maps after subtraction of the best-fit model, no residual emission is detected above 3$\sigma$.
To increase sensitivity, especially to extended emission, we perform a stacking analysis of the model-subtracted visibilities for nine compact sources, using the {\tt STACKER} tool \citep{2015MNRAS.446.3502L}. 
The phase center is shifted to the center position of the best-fit model before the stacking. 
A clean map is created from the stacked visibility with $uv$-taper of the on-sky FWHM=1\arcsec.0 and the resultant synthesized beam size is 0\arcsec.81$\times$0\arcsec.87. 
The residual emission is detected at 4.3$\sigma$ and its flux density within 2\arcsec.0 aperture is $S_\mathrm{extra}$=0.42 mJy, corresponding to 21\% of the total average flux. 

Conservatively assuming that this residual flux originates outside the half-light radius, we calculate the corrected half-light radius, $R_{1/2,\mathrm{cor}}$, which encloses half of the total flux, $S_{1/2,\mathrm{cor}}=(S_\mathrm{model}+S_\mathrm{extra})/2$, in the primary exponential component. 
The amount of correction depends on the ratio of $S_{1/2,\mathrm{cor}}/S_\mathrm{model}$.
This has the largest impact on size measurements for U4-32351 with the faintest model flux as $R_{1/2,\mathrm{cor}}$ corresponds to a radius enclosing 78\% of the flux in the exponential model.

\begin{figure}[t]
\begin{center}
\includegraphics[scale=1.]{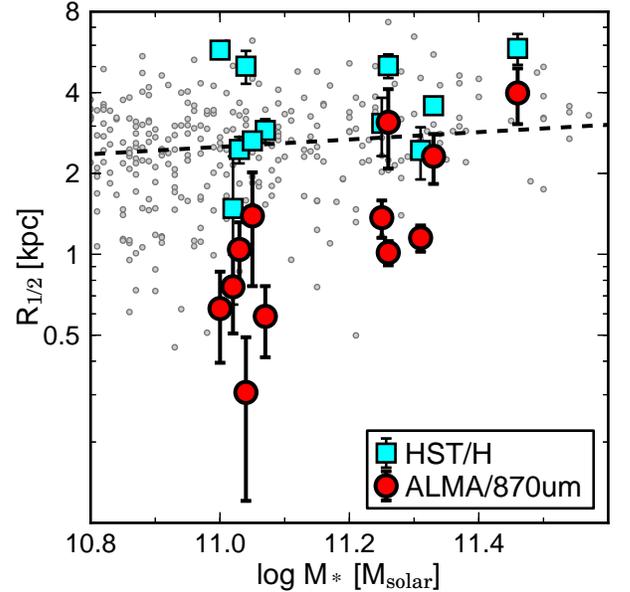}
\caption{Comparison of circularized half-light radii at ALMA/870 $\mu$m (red circles) with those at HST/$H_{160}-$band (cyan squares) for our sample of 12 galaxies with size measurements at 870 $\mu$m. 
Gray circles indicate the rest-optical sizes for star-forming galaxies around the main-sequence at $z=1.9-2.7$, which are drawn from the 3D-HST survey. 
A dashed line shows their fitting function, $\log(R_{1/2})=0.14\log(M_*)-1.11$.
 \label{fig;size}}
\end{center}
\end{figure}

For nine out of the 12 star-forming galaxies, the corrected 870 $\mu$m sizes are less than 1.5 kpc (Figure \ref{fig;size}, Table \ref{tab;2}), which is more than a factor of 2 smaller than their rest-optical sizes and is comparable with optical sizes of massive quiescent galaxies \citep[e.g.,][]{2007MNRAS.382..109T,2007ApJ...671..285T,2008ApJ...677L...5V,2012ApJ...746..162N}.
They have an extended exponential profile with $R_{1/2,1.6\mu\mathrm{m}}$=3.2 (1.5--5.8) kpc and S$\acute{\mathrm{e}}$rsic index $n$=1.2 (0.5--1.9) in the rest-optical maps. 
In the stellar mass range of log($M_*/M_\odot)<11$, star-forming galaxies could form stars within somewhat larger disks than the bulk of stars to slowly grow in size with increasing stellar mass as seen in the mass-size relation of normal star-forming galaxies \citep{2016ApJ...828...27N,2016arXiv160707710R}. 
Our best ALMA sample of 12 star-forming galaxies is all massive with log($M_*/M_\odot)>11$.
Their individual detection of compact dust emission above the Schechter mass suggests that star formation preferentially occurs in the compact central region.
This has a potential to change galaxy morphologies from disk-dominated to bulge-dominated with high stellar mass surface densities (see next section).

In the analysis of size measurements, we do not include two massive star-forming galaxies with log($M_*/M_\odot)>11$ in the parent sample of galaxies identified by the narrow-band survey.
One is not observed with ALMA and the other one (U4-36568) is not detected in the high-resolution map (Figure \ref{fig;MS}).
Given the high completeness of 86\% (12/14) in the stellar mass range, our results are not significantly affected by the sample selection.
Therefore, we find massive galaxies to commonly form stars in the extremely compact central region as at least 64\% (9/14) have small 870 $\mu$m sizes of $R_{1/2,870\mu\mathrm{m}}<$1.5 kpc.
This result is in excellent agreement with similar and independent evidence coming from an ALMA/870 $\mu$m study of 6 massive star-forming galaxies at $z\sim2$ \citep{2016ApJ...827L..32B}. 
\cite{2016ApJ...827L..32B} find that the mean half-SFR radius is $\sim$30\% smaller than the mean half-mass radius.
The main difference between our work and \cite{2016ApJ...827L..32B} is that they pre-select only optically compact star-forming galaxies while our study almost completely select main-sequence galaxies.


\section{Bulge formation in extended, rotating disks}
\label{sec;bulge}

Given that our ALMA sample is already massive, $\log(M_*/M_\odot)>11.0$, they are likely to soon thereafter quench the active star formation and to be observed as quiescent galaxies in the local Universe. 
Quiescent galaxies are always smaller than star-forming galaxies at any redshift and any stellar mass and have a cusp profile ($n>2$) unlike star-forming galaxies with exponential disks \citep{2011ApJ...742...96W,2014ApJ...788...28V}. 
A spatial distribution of stars within galaxies would not be changed unless a violent process like major mergers happens. 
However, centrally-concentrated star formation reduces the half-light or half-stellar-mass radii of galaxies and their S$\acute{\mathrm{e}}$rsic index would increase by central bulge formation.

We quantitatively assess the possibility of bulge formation in our sample of the 12 massive galaxies with reliable size measurements of dust continuum emission. 
Quiescent galaxies generally have a dense core with high stellar mass surface densities within 1 kpc of galaxy centers of $\log(\Sigma M_\mathrm{*,\mathrm{1kpc}}/M_\odot~\mathrm{kpc}^{-2})=10$ while star-forming galaxies mostly do not \citep{2014ApJ...791...45V,2015arXiv150900469B}. 
For our sample, we create stellar mass maps by spatially resolved SED modeling with multi-band HST data \citep{2012ApJ...753..114W, 2014ApJ...788...11L} to calculate stellar mass surface densities within 1 kpc from the 870 $\mu$m center. 
None of our sample satisfy the criterion of a dense core at the current moment (Table \ref{tab;2}).
The spatial distribution of star formation within galaxies allows us to understand when the dense core is formed by subsequent star formation.
Exploiting the geometric information of the best-fit exponential models at 870 $\mu$m, we derive the SFR surface densities within the central 1 kpc ($\Sigma$SFR$_\mathrm{1kpc}$) from the Spitzer/Herschel-based total SFRs over galaxies.
For nine galaxies with compact dust emission of $R_{1/2,870\mu\mathrm{m}}<1.5$ kpc, they are intensely forming stars in the central region with $\Sigma$SFR$_\mathrm{1kpc}$=40 (19--65) $M_\odot$yr$^{-1}$kpc$^{-2}$ (Table \ref{tab;2}). 
Then, bulge formation timescales to reach $\log(\Sigma M_\mathrm{*,\mathrm{1kpc}}/M_\odot$kpc$^{-2})=10$ are estimated by

\begin{equation}
\tau_\mathrm{bulge}=\frac{10^{10}-\Sigma M_{*,\mathrm{1kpc}}}{w\times \Sigma \mathrm{SFR}_\mathrm{1kpc}},
\label{eq;1}
\end{equation}

\noindent
taking into account mass loss due to stellar winds ($w=0.6$ in Chabrier initial mass function, see also \citealt{2014ApJ...791...45V}). 
The estimated bulge formation timescales are $\langle\log \tau_\mathrm{bulge}\rangle= 8.47$ (8.16--8.79) for the nine galaxies with $R_{1/2,870\mu\mathrm{m}}<1.5$ kpc.
They can complete the dense core formation by $z=2$ when the current level of star formation is maintained for several hundred Myr. 
Galaxies forming stars in disks as extended as the rest-optical light would have to keep the current star formation for a longer time ($\sim$2 Gyr). 
This is not consistent with stellar populations obtained in high-redshift quiescent galaxies, where timescales for star formation are $\tau<1$ Gyr \citep[e.g.,][]{2015ApJ...799..206B, 2015ApJ...808..161O}. 

We also estimate gas depletion timescales for our ALMA sample using the \cite{2015ApJ...800...20G} scaling relations, combining CO-based, Herschel far-infrared-dust based and submillimeter-dust based estimates, in order to average over the systematic uncertainties inherently present in all of these techniques.
We use the updated version of this scaling relation (Tacconi et al. in prep), $\log(M_\mathrm{gas}/\mathrm{SFR})=0.15-0.79\log(1+z)-0.43\log(\mathrm{sSFR}/\mathrm{sSFR}_{\mathrm{MS}})+0.06(\log M_*-10.5)$ where $\mathrm{sSFR}_{\mathrm{MS}}$ is the specific star formation rate on the main-sequence line of \cite{2014ApJ...795..104W} at given redshift and stellar mass.
We adopt uncertainties of $\pm$0.24 dex for the $\log(M_\mathrm{gas}/\mathrm{SFR})$ \citep{2015ApJ...800...20G}.
The gas is partly consumed by star formation and partly ejected by outflows from the central region with comparable rates to SFR, $\eta\times$SFR ($\eta\sim1$), especially for massive galaxies \citep{2014ApJ...796....7G}. 
Thus, gas depletion timescales are re-defined as

\begin{equation}
\tau_\mathrm{depl}=\frac{M_\mathrm{gas}}{\mathrm{SFR}(1+\eta)}.
\label{eq;2}
\end{equation}

\noindent
The gas depletion timescales are, on average, similar to the bulge formation timescales, $\langle\tau_\mathrm{bulge}/\tau_\mathrm{depl}\rangle\sim1.2$ for the nine galaxies with $R_{1/2,870\mu\mathrm{m}}<1.5$ kpc, suggesting that the formation of a dense core does not necessarily require additional gas accretion onto the galaxies.

\begin{figure}
\begin{center}
\includegraphics[scale=1.]{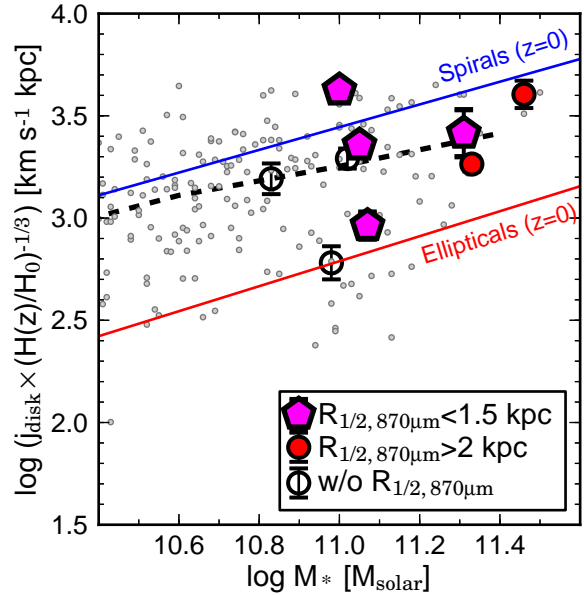}
\caption{Specific angular momentum of disks versus stellar mass for our ALMA sample. 
Magenta pentagons and red circles denote galaxies with $R_{1/2,870\mu\mathrm{m}}<1.5$ kpc and with $R_{1/2,870\mu\mathrm{m}}>2.0$ kpc, respectively.
The kinematic properties are derived from ionized gas.
Gray circles indicate the KMOS$\mathrm{3D}$ sample of galaxies at $z=0.8-2.6$ along with their median values in stellar mass bins of 0.4 dex (dashed line). 
The relations for local spiral and elliptical galaxies are shown by a blue and red solid line, respectively \citep{2013ApJ...769L..26F}.
Here, the redshift dependence is removed by multiplying $j_\mathrm{disk}$ with $H(z)^{1/3}$.}
\label{fig;AM}
\end{center}
\end{figure}

Next, we look at the kinematic properties for nine galaxies that were observed as part of the KMOS$^\mathrm{3D}$ program.
Six out of them have 870 $\mu$m size measurements (Table \ref{tab;1}).
We note that they are all rotation-supported ($v_{rot}/\sigma_0>3$).
Figure \ref{fig;AM} shows specific angular momenta as a function of stellar mass for galaxies at $z=0.8-2.6$ from the KMOS$^\mathrm{3D}$ survey \citep{2016ApJ...826..214B}.
They span a range of disk angular momenta from local spirals to ellipticals \citep{2013ApJ...769L..26F}.
A lower offset at fixed stellar masses suggests that galaxies have lost a significant fraction of their original angular momentum (e.g., major mergers \citealt{2014MNRAS.444.3357N,2015ApJ...804L..40G}) or that they had a small initial angular momentum.
We find the specific angular momentum of galaxies with $R_{1/2,870\mu\mathrm{m}}<1.5$ kpc to be broadly consistent with a large sample of primarily mass-selected galaxies from the KMOS$^\mathrm{3D}$ survey.
Our result plausibly indicates that these galaxies as a group are not all galaxies with very low angular momentum, either due to large angular momentum loss of the baryonic component or due to a small initial dark matter angular momentum parameter. 
The compact nuclear dust components we have detected are most likely caused by internal angular momentum redistribution, such as has been proposed by recent observations and theoretical studies \citep{2016ApJ...826..214B, 2015MNRAS.450.2327Z, 2014MNRAS.438.1870D}.

Finally, we speculate that the halo masses inferred from our KMOS observations and a Monte-Carlo modeling are log($M_\mathrm{halo}/M_\odot)>$12 \citep{2016ApJ...826..214B}. 
In such massive halos, infalling gas along cosmic filaments is heated to the halo virial temperature by shocks and cold gas is not directly supplied to galaxies \citep[e.g.,][]{2006MNRAS.368....2D}. 
Given the bulge formation timescales are comparable with the gas depletion timescales by central starbursts and outflows, they can naturally quench star formation soon after the dense core is formed. 
Even if some amount of cold gas accretes onto galaxy disks after cooling, a steep potential by the dense core (morphological quenching) helps galaxies to keep quiescent properties after nuclear starbursts consume all central gas or outflows eject it. 
Therefore, galaxies with compact dust emission would be a key population for understanding the morphological and star formation evolution from star-forming disks to quiescent spheroids at the massive end of the main-sequence.

On the other hand, our observations detect relatively extended dust emission of $R_{1/2,870\mu\mathrm{m}}>2$ kpc from the remaining three massive galaxies. 
Although the bulge formation timescale is longer than the gas depletion timescale, $\tau_\mathrm{bulge}/\tau_\mathrm{depl}>2$, two of them show a high S$\acute{\mathrm{e}}$rsic index, $n>2$, in the rest-optical, suggesting the bulge is already formed. 
They can directly become large quiescent galaxies after consuming gas, not through the compaction phase \citep{2015ApJ...813...23V}. 
This mode would become dominant at a later epoch when the number density of optically compact galaxies decreases \citep{2013ApJ...765..104B}.

\section{Summary}

We have presented 0\arcsec.2-resolution 870 $\mu$m observations for 25 H$\alpha$-selected star-forming galaxies on/around the main sequence at $z=2.2$ and $z=2.5$ with ALMA. 
We have robustly detected the dust continuum emission from 16 galaxies and measured the half-light radii for the best sample of 12 massive galaxies with $\log (M_*/M_\odot)>11$.
In this paper, we have investigated dense core formation in extended star-forming disks and verified the evolutionary scenarios from disk-dominated galaxies to bulge-dominated ones.

\begin{enumerate}

\item We have discovered nine massive galaxies associated with extremely compact dust emission with $R_{1/2,870\mu\mathrm{m}}<1.5$ kpc.
In spite of the compact appearance at 870 $\mu$m, they have an extended, rotating disk with $R_{1/2,1.6\mu\mathrm{m}}=3.2$ kpc and $n_{1.6\mu\mathrm{m}}=1.2$ in the rest-optical.
The difference of morphologies between dusty star formation and stars suggests they would reduce the half-light or half-mass radius by the subsequent star formation and increase the S$\acute{\mathrm{e}}$rsic index.
Given the high completeness in the stellar mass range of $\log (M_*/M_\odot)>11$,
they are likely a common population of massive star-forming galaxies at $z\sim2$.

\item Galaxies with $R_{1/2,870\mu\mathrm{m}}<1.5$ kpc can complete the formation of a dense core in several hundred Myr if the current level of star formation is maintained.
This would be reasonable because the bulge formation timescales are comparable with the gas depletion timescales by star formation and nuclear outflows.
Therefore, they can naturally quench star formation after the dense core is formed.

\item Three massive star-forming galaxies show somewhat extended dust emission with $R_{1/2,870\mu\mathrm{m}}>2.0$ kpc.
As two of them already have a cusp profile ($n>2$) rather than exponential disks, they can evolve into extended quiescent galaxies.
This direct pathway is not the norm at $z\sim2$, but could dominate at later epochs.

\item For our ALMA sample, available integral field observations of \ha emission with KMOS provide the kinematic parameters of ionized gas such as rotation velocity, local velocity dispersion, and specific angular momentum. 
They are all rotation-supported disks and their disk angular momenta are consistent with a large sample of mass-selected star-forming galaxies at $z=0.8-2.6$ in the KMOS$^\mathrm{3D}$ survey.
Our finding suggests that internal processes are primarily responsible for the bulge formation rather than major mergers.

\end{enumerate}

\

We thank the anonymous referee who gave us many useful comments, which improved the paper.
This paper makes use of the following ALMA data: ADS/JAO.ALMA\#2012.1.00245.S and 2013.1.00566.S. ALMA is a partnership of ESO (representing its member states), NSF (USA) and NINS (Japan), together with NRC (Canada), NSC and ASIAA (Taiwan), and KASI (Republic of Korea), in cooperation with the Republic of Chile. 
The Joint ALMA Observatory is operated by ESO, AUI/NRAO and NAOJ. 
We thank the staff at Paranal Observatory for their helpful support. Data analysis was in part carried out on the common use data analysis computer system at the Astronomy Data Center, ADC, of the National Astronomical Observatory of Japan. K.T. was supported by the ALMA Japan Research Grant of NAOJ Chile Observatory, NAOJ-ALMA-34.
This paper is produced as a part of our collaborations through the joint project supported by JSPS and DAAD under the Japan - German Research Cooperative Program.
S I. acknowledges the support of the Netherlands Organization for Scientific Research (NWO) through the Top Grant Project 614.001.403.

\bibliographystyle{apj}
\bibliography{tadaki_2016}

\end{document}